\documentclass[9pt,twocolumn,twoside,lineno]{npj-new}

\usepackage[super, numbers, sort&compress]{natbib}
\usepackage{hyperref}
\usepackage[T1]{fontenc}
\usepackage{tgtermes}
\usepackage{mathrsfs}
\usepackage{float}
\usepackage{xcolor}
\usepackage{subcaption}
\usepackage{dcolumn}
\usepackage{bm}
\usepackage{multirow}
\usepackage{array}
\usepackage{siunitx} 
\usepackage{booktabs}
\usepackage{calligra}
\definecolor{myblue}{RGB}{0,93,133}
\definecolor{myorange}{RGB}{221,107,0}
\DeclareMathAlphabet{\mathcalligra}{T1}{calligra}{m}{n}
\DeclareFontShape{T1}{calligra}{m}{n}{<->s*[2.2]callig15}{}

\newcolumntype{P}[1]{>{\centering\arraybackslash}p{#1}}

\AtBeginDocument{
\heavyrulewidth=.08em
\lightrulewidth=.05em
\cmidrulewidth=.03em
\belowrulesep=.65ex
\belowbottomsep=0pt
\aboverulesep=.4ex
\abovetopsep=0pt
\cmidrulesep=\doublerulesep
\cmidrulekern=.5em
\defaultaddspace=.5em
}

\templatetype{npjresearcharticle} 

\title{Opinion dynamics on social proximity disordered networks}

\author[1,2]{Fellipe Aranha}
\author[1,3]{Giuliano G. Porci\'uncula}
\author[4]{Adauto J. F. de Souza}
\author[5]{Paulo R. A. Campos}
\author[1,5]{Andr\'e L. M. Vilela}

\affil[1]{\textit{F\'isica de Materiais, Universidade de Pernambuco, Recife, PE 50720-001, Brazil}}
\affil[2]{\textit{Department of Physics, Bar-Ilan University, Ramat-Gan, 52900, Israel}}
\affil[3]{\textit{Department of Physics, Northeastern University, Boston, MA 02115, USA}}
\affil[4]{\textit{Departamento de Física, Universidade Federal Rural de Pernambuco, Recife, PE 52171-900, Brazil}}
\affil[5]{\textit{Departamento de F\'isica, Universidade Federal de Pernambuco, Recife, PE 50670-901, Brazil}}

\leadauthor{L. F. Aranha de Oliveira} 

\onecolumn
\begin{abstract}
Understanding how local social pressure shapes collective opinion formation is essential in modern society, with implications in sociology, politics, finance, and technology. Complex networks are a powerful framework for investigating these processes and for representing social interactions in emergent phenomena, and canonical models such as random, scale-free, and small-world networks remain most relevant when Euclidean geometric proximity and contiguous interactions are not the dominant organizing factors. Our investigation explores the effects of Gabriel graphs, a spatially embedded network that utilizes proximity-based interactions, on social opinion dynamics. Using a geometric disorder parameter that randomly displaces agent coordinates, we transition from square lattices of social interactions to spatial random networks that preserve local constraints while introducing strong topological heterogeneity. Opinion dynamics operate through social pressure, with a level of nonconformity that affects dissent among opinions. We analyze magnetization, susceptibility, and Binder's fourth-order cumulant to characterize the emergence and breakdown of global social consensus. The model exhibits second‑order phase transitions whose behavior depends on both geometric disorder and nonconformity levels. We use volumetric scaling and provide a reference frame appropriate for characterizing criticality in regular and heterogeneous interaction topologies. We find that the standard critical exponents remain compatible with the two‑dimensional Ising universality class. Furthermore, the results suggest that spatially constrained, planar, and globally connected interaction topologies preserve the dimensionality of space in critical dynamics analysis.
\end{abstract}

\dates{\today}

\begin{document}

\maketitle
\thispagestyle{firststyle}
\ifthenelse{\boolean{shortarticle}}{\ifthenelse{\boolean{singlecolumn}}{\abscontentformatted}{\abscontent}}{}


\twocolumn

\dropcap{C}omplex systems are a central focus of contemporary physics, with the primary objective of understanding how local interactions among many interconnected entities generate system-wide organization, spontaneous order, and collective phenomena, including riots, epidemics, financial crises, and consensus formation, among others \cite{2015HenriRiots, 2005Sporns_The, 2018Bassett_On, 2000Jeong_The, 2005Guimera_Functional, 2017Yilmaz_Metabolic, 2009Castellano_Statistical, Raoufi2025Messengers, Lipatov2025Propaganda, 2001PastorSatorras_Epidemic, 2017Wang_Unification, 2024DeArruda_Contagion, 1999Mantegna_Hierarchical, 2020Aslam_Network}. Instead of explaining systems solely in terms of their individual components, complex system science emphasizes the emergence of macroscopic behavior. Using analytical, computational, and stochastic approaches, scientists investigate nonlinear dynamics, self-organization, and critical phenomena, with applications spanning physics, biology, chemistry, engineering, sociology, and economics. A natural framework for studying such systems is network science, which models relations between discrete entities and quantifies the structure and strength of their associations \cite{2009Mitchell_Complexity, 2020BattistonNetworks}.

The complex network theory equips researchers with tools to model and investigate various real-world phenomena across disciplines driven by interconnectivity, from synaptic interactions in the brain \cite{2005Sporns_The, 2018Bassett_On} and nutrient synthesis in the body \cite{2000Jeong_The,2005Guimera_Functional,2017Yilmaz_Metabolic} to opinion dynamics in society \cite{2009Castellano_Statistical, Raoufi2025Messengers, Lipatov2025Propaganda}, disease transmission \cite{2001PastorSatorras_Epidemic, 2017Wang_Unification, 2024DeArruda_Contagion}, and market dynamics in stock exchanges \cite{1999Mantegna_Hierarchical, 2020Aslam_Network}. Most social, biological, and technological networks exhibit nontrivial interaction structures, with connectivity patterns that are neither purely regular nor purely random. Such properties include heterogeneous connectivity distributions, high clustering among components, and small average path length and diameter, properties that control how efficiently information, influence, and dependencies spread between agents or components in the network \cite{2016barabasi_network, 2024Artime_Robustness}.
 
An underlying interaction structure is fundamental in shaping collective behavior, and canonical network models, such as random, scale-free, and small-world networks, have been extensively employed to describe and study the intricate relationships in systems with numerous constituents \cite{2016barabasi_network,2024Artime_Robustness}. These models are inherently non-planar due to their high dimensionality, allowing links without geometric constraints. In addition, the structural characterization of such networks relies on key topological properties, including measures of degree distribution $p(k)$, which captures the connectivity patterns among components; average path length $\langle \ell \rangle$, which quantifies the typical shortest-path distance between node pairs; and average clustering coefficient $\langle C \rangle$, which reflects the tendency of the network to form cohesive groupings. In particular, small-world, random, and scale-free networks exhibit the small-world property, in which $\langle \ell \rangle$ remains relatively small compared to the system size, demonstrating that long-range associations enable networks to achieve short-scale properties. In the social context, these models capture nonlocal interactions satisfactorily, as observed in online communication, social media platforms, and general information exchange systems.

Regular lattice models provide the simplest representation of spatially constrained interactions and have played a central role in statistical physics and opinion dynamics. However, many real systems exhibit substantial structural heterogeneity while preserving local spatial limitations. In contrast to interaction networks that permit arbitrary long-range connections, proximity-based graph models provide a realistic representation of the interaction structure when spatial locality constrains connectivity. Examples include face-to-face social interactions, workplace interaction networks, wireless sensor and mobile communication networks, ecological dispersal networks, regional communication, cultural diffusion networks, and similar phenomena \cite{2019RossiOn, 2021FrysztackiThe, 2022Barthelemy_Spatial}. In such systems, geometric proximity limits connections and yields distinct structural properties, including planarity, bounded degree, and longer average path lengths. In this context, the foregoing complex network models lose their descriptive power, as they permit long-range links, since Euclidean distances are irrelevant in connection rules.

To capture the spatial organization present in real-world systems, researchers have developed several proximity-based network models, including random geometric graphs, Delaunay triangulations, Voronoi graphs, Gabriel graphs, and relative neighborhood graphs \cite{2007Nekovee_Worm, 2022Barthelemy_Spatial, 1961Gilbert_Random}. In these systems, structural properties do not arise independently but emerge from the spatial arrangement of nodes and the interplay between Euclidean distance and the proximity definition adopted. For instance, random geometric graphs do not restrict crossing edges, although the Euclidean distance between pairs of nodes within a fixed neighborhood radius explicitly determines connectivity \cite{1961Gilbert_Random}. Alternatively, exclusion-area graphs define proximity through the emptiness of a spatial region. In these models, the proximity rule establishes a connection only if no other node lies within a specific interaction region, naturally promoting planarity.

Gabriel graphs networks connect two nodes only if no third node lies inside the disk whose diameter is the segment joining the pair \cite{gabriel1969new, matula1980properties}. This geometric rule produces sparse planar networks with localized, distance-dependent interactions, making Gabriel graphs especially suitable for systems with relevant local influence and efficient information propagation. Moreover, Gabriel graphs are subgraphs of the Delaunay triangulation and still contain the minimum spanning tree. As a consequence, these networks provide sufficient yet not excessive interconnectivity among neighboring nodes, naturally balancing local cohesion and spatial sparsity. Gabriel graphs are a natural choice for investigating wireless communication networks \cite{2002Li_Sparse, 2007Wan_On}, geographic and routing systems \cite{2013Keller_Spatial}, and transport processes \cite{2014Cetinkaya_A}, in which connectivity emerges from spatial proximity, thereby capturing the physical and infrastructural constraints of real-world socioeconomic systems. Such characteristics make these graphs especially suitable for investigating how spatial constraints on social interactions give rise to large-scale collective phenomena.

A fundamental question in opinion dynamics concerns how microscopic social interactions give rise to macroscopic collective phenomena, including cultural dissemination, social influence, consensus formation, polarization, fragmentation, and nonequilibrium phase transitions. One way to analyze these phenomena is through similarity-based models, in which interactions depend on cultural overlap or opinion proximity. The Axelrod model is a paradigmatic example, since it shows that local convergence does not necessarily destroy global diversity \cite{Axelrod1997Culture}. In this model, agents possess several cultural features, each one with different traits, and neighboring agents interact only when they share at least one trait. Thus, increasing the number of features can promote partial overlap and convergence, whereas increasing the number of traits reduces similarity, suppresses interaction, and allows monocultural regions to persist. Similarly, the Deffuant--Weisbuch and Hegselmann--Krause models describe bounded-confidence communication, in which agents update their continuous opinions only when their differences fall within a confidence threshold, leading to either consensus or fragmentation \cite{2000DeffuantMixing, 2002RainerHKmodel}. These models show how selective interaction based on likeness can simultaneously promote local agreement and sustain global diversity.

Social-pressure models, in which agents tend to align with dominant or persuasive local groups, deliver a complementary approach to modeling opinion formation. Such sociophysical approaches highlight how large-scale collective organization, consensus formation, polarization, and social fragmentation emerge from the competition between local alignment, persuasive social influence, and stochastic disagreement. In this context, the majority-vote model captures the competition between social conformity and individual nonconformity, leading to a nonequilibrium transition between ordered and disordered social states \cite{1992Oliveira_Isotropic}. Related mechanisms appear in the Galam model, where agents interact in small, randomly formed groups and adopt the local majority opinion, and in the Sznajd model, where aligned pairs persuade their neighbors through an outward flow of information \cite{Galam2008Sociophysics, SznajdWeron2021Review}. The $Q$ voter model extends this idea by assuming that a group of neighbors can collectively influence a target agent, usually when they share the same opinion \cite{Castellano2009QVoter}.

The majority-vote model with noise captures social opinion formation in systems with two competing alternatives, such as choosing between two different political parties, deciding whether to buy or sell a stock, or even accepting or rejecting vaccination programs. Analogous to the role of the Ising model in magnetic systems, the majority-vote model provides a minimal representation of social interactions that gives rise to a wide variety of emergent phenomena, including nonequilibrium phase transitions and well-defined critical behavior. In this model, agents interact through social connections and tend to align with their neighbors' majority opinion with probability $1-q$, while adopting the opposite opinion with probability $q$, where $q$ denotes the population's nonconformity level. On square lattices of social interactions, agents interact with exactly four neighbors, and the model exhibits a continuous order-disorder phase transition at a critical level of nonconformity $q_c \approx 0.075$, and shares the same universality class as the two-dimensional ferromagnetic Ising model \cite{1992Oliveira_Isotropic}. 

\begin{figure*}[!ht]
  \centering
  \includegraphics[width=1.0\linewidth]{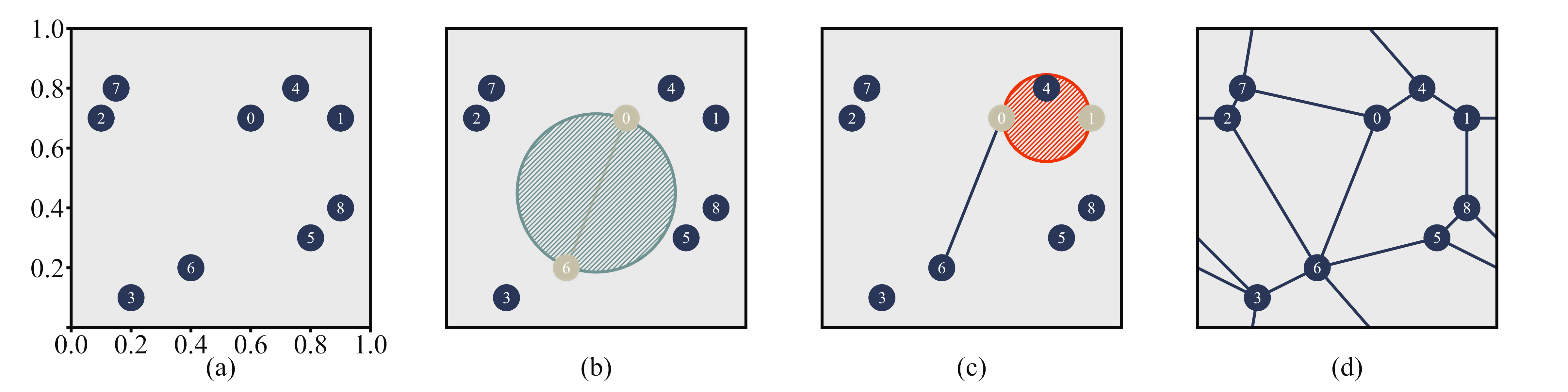}
\caption{\textbf{Illustration of Gabriel graph construction from spatially distributed agents.} (a) Initial spatial distribution of agents embedded in a two-dimensional $1 \times 1$ domain. (b) Candidate connection between nodes $0$ and $6$. The hatched green circle represents the disk whose diameter is defined by the candidate edge. Since no other node lies inside the disk, the connection satisfies the Gabriel condition and is accepted. (c) Candidate connection between nodes $0$ and $1$. In this case, node $4$ lies inside the corresponding disk, denoted by the hatched red circle, violating the Gabriel condition and preventing edge formation. (d) Gabriel graph obtained after testing all node pairs according to the geometric proximity criterion. The construction naturally generates planar proximity networks with spatially constrained interactions and heterogeneous local connectivity patterns.}  \label{fig:constructGG}
\end{figure*}

Subsequent investigations extended the majority-vote opinion dynamics using distinct social interaction structures, including small-world networks with long-range shortcuts that promote social ordering against $q$, random graphs of social contacts, yielding expanding collective agreement with increasing average connectivity, scale-free networks capturing the presence of highly connected agents and their function to consensus robustness, and others \cite{2003CamposSmallWorld, 2005PereiraMajorityVote, 2006LuzMajorityVote, 2010WuMajorityVote}. Such applications demonstrate how distinct interaction structures shape emergent collective phenomena, providing successful descriptions of order-disorder transitions, rich phase-diagram behavior, and diverse universality classes.

Recent studies have extended the majority vote rules to account for additional social mechanisms. These works introduce agent heterogeneity and noise heterogeneity, pushing the framework away from an idealized homogeneous society \cite{2018VilelaEffect, 2018KrawieckiSpinGlass, 2024OliveiraEntropy, 2016VieiraPhase}; history dependence, showing how past decisions constrain present responses \cite{2017HarunariPartial, 2018EncinasFundamental, 2020ChenNonMarkovian}; multi-state extensions, revealing how consensus emerges from a richer landscape of competing alternatives \cite{Melo2010ThreeStateMV, 2020VilelaThree, 2022ZubillagaThree}; and visibility effects \cite{Vilela2021LimitedVisibility, 2025porciuncula_Consensus}, illustrating how synthetic neighborhoods reshape opinion perception and alter consensus formation. The model has also inspired new opinion-dynamics frameworks for economic systems and financial markets, in which herding among traders, market uncertainty, and strategic opposition generate macroscopic phenomena such as volatility clustering, heavy-tailed return distributions, and long-range correlations \cite{2019VilelaMajorityVote,2022GranhaOpinion,2022ZubillagaA}. These developments validate the adaptability of majority-vote rules for linking simple local interaction rules to a wide range of emergent collective behaviors.

The body of research on the majority-vote model focuses on various dynamical features and interaction neighborhoods, while neglecting spatial constraints. The spatial embedding approach to social interactions offers a novel perspective on how local social influence, individual nonconformity, and geometrically constrained associations shape collective opinion dynamics. In this context, we explore how Gabriel graphs with Gaussian positional disorder influence opinion evolution, consensus formation, and nonequilibrium critical behavior employing the majority-vote dynamics. Specifically, we investigate the relation between two competing mechanisms: the positional disorder parameter $s$, which modifies the structural properties of Gabriel graphs, and the social nonconformity parameter $q$, which controls stochastic opinion deviations from local majority alignment. Using Monte Carlo simulations, we examine how spatial disorder alters connectivity patterns and shifts the critical noise associated with the consensus--dissent transition. Our approach provides a practical framework for investigating opinion dynamics in geographically distributed societies, where heterogeneous interaction structures strongly influence social behavior. In the following sections, we present the observables and metrics, outline the simulation setup and computational details, and discuss the results.

\section*{Methods} \label{sec:methods}
\subsection*{Modeling and research design}

We consider a system of $N$ agents, indexed by $i \in \{1, \ldots, N\}$, each of which has an opinion variable that can assume one of two states $\sigma_i = +1$ or $-1$ and evolves in discrete time steps. Initially, we arrange each agent on a regular square lattice and embedded in a two-dimensional domain, occupying stationary positions $\vec{r}_i = x_i \hat x + y_i \hat y \in \Gamma$, where $\Gamma \subset \mathbb{R}^2$ is a unitary square region with periodic boundary conditions. A parameter $s$ introduces geometric disorder by random displacements of the lattice positions, and proximity rules of spatial networks constrain social opinion-influence groups of each agent. We measure time discretely in Monte Carlo steps (MCS), using asynchronous random evolution, where one unit of time, $t \to t+1$, corresponds to $N$ attempted updates of spin variables, each selected uniformly at random. Within this framework, Gabriel graphs determine connectivity, and opinion dynamics follow the majority-vote rule with a nonconformity parameter $q$ that enables stochastic deviations from local consensus.

We summarize the following model and design components to capture the spatial and informational constraints driving opinion dynamics and collective emergent critical phenomena:
\begin{itemize}
	\item \textbf{Social proximity structure.} Agents continuously exchange information with their local neighbors through a proximity-based interaction network represented by Gabriel graphs. A disorder parameter $s$ controls random perturbations of agent positions, allowing the interaction structure to continuously transition from regular lattice configurations to spatially heterogeneous proximity networks.
	\item \textbf{Local consensus formation.} Individual opinion updates depend on the states of neighboring agents connected through the social network, capturing how local majority influence propagates through spatially constrained interactions. Although decisions depend on local information, neighbor-to-neighbor interactions enable the emergence of large-scale collective phenomena.
	\item \textbf{Nonconformity-driven transition.} In the absence of nonconformity, $q = 0$, agents always adopt the local majority opinion, leading to global consensus with maximum social order. As $q$ increases, stochastic deviations from local majority alignment progressively weaken collective order. Beyond a critical noise level $q_c$, the system undergoes an order-disorder phase transition characterized by the loss of global consensus.
\end{itemize}

In the following sections, we present the observables and metrics, outline the simulation setup and computational details, and discuss the resulting collective phenomena and critical behavior.

\begin{figure*}[!ht]
	\centering
\includegraphics[width=1.0\linewidth]{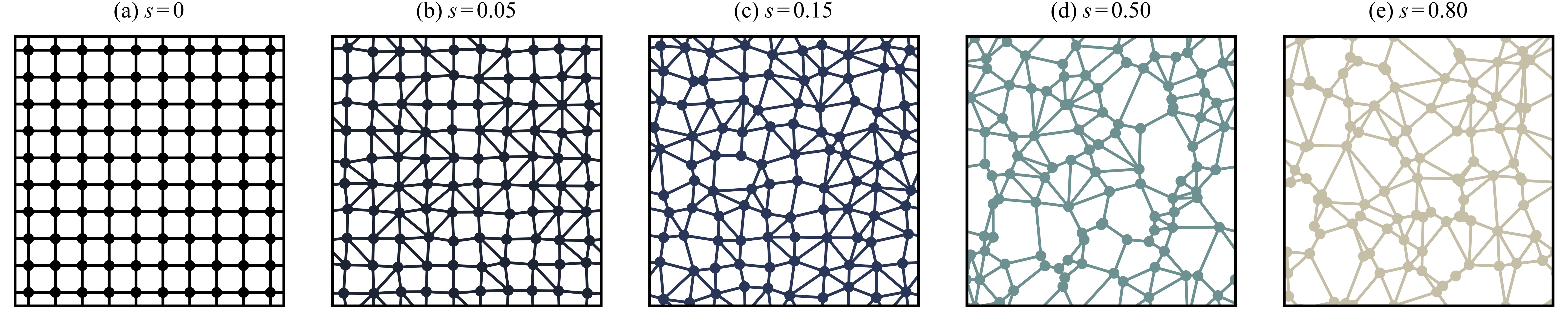}
	\caption{\textbf{Structural evolution of Gabriel graph social networks under increasing spatial disorder.} Increasing disorder progressively breaks the lattice's geometric regularity, alters local connectivity patterns, and generates heterogeneous proximity-based interaction structures. Representative realizations of the interaction networks for different disorder levels $s$: (a) $s = 0$, corresponding to the regular square-lattice regime; (b) $s = 0.05$, where weak positional perturbations introduce local diagonal connections; (c) $s = 0.15$, associated with the connectivity-enhanced regime and near maximum average coordination; (d) $s = 0.50$, where the network becomes spatially heterogeneous with irregular local neighborhoods; and (e) $s = 0.80$, representing the strongly disordered regime of spatial randomness.} 
	\label{fig:disorder_snet}
\end{figure*}

\subsection*{Gabriel graph social networks with variable disorder}

Social networks provide the interaction structure through which information, opinions, and influence propagate among individuals. When spatial proximity constrains social associations, geometric considerations naturally shape the structure of connectivity and information propagation. In such settings, proximity graphs provide a formal representation for encoding local connectivity solely from spatial relationships rather than by topological rules. Among different proximity graphs, Gabriel graphs offer a particularly suitable description, as they capture nearest-neighbor interactions while preserving sparsity and planarity. In such networks, nodes $i$ and $j$ connect if they satisfy the inequality

\begin{equation}\label{eq:gg}
d^2_{ij} < d^2_{ik} + d^2_{jk}, \; \forall k \in \{1, \ldots, N\},\; k \neq i,j.
\end{equation}
This condition requires that, for every other node $k$, the pair $(i,j)$ satisfies the inequality above. Geometrically, this is equivalent to expressing that two individuals interact only if no third agent lies within the disk whose diameter is the segment connecting $i$ and $j$. This condition suppresses long-range crossings and redundant local links, generating interaction networks without crossing edges. Consequently, the resulting topology reflects the local geometric organization of agents and naturally incorporates spatial constraints into the social interaction structure.

Fig.\ref{fig:constructGG} illustrates the geometric criterion underlying the construction of Gabriel graph social networks. In panel (a), agents are distributed in a two-dimensional spatial domain, where each point represents an agent in the social system. Applying the proximity condition of Eq.(\ref{eq:gg}) to every pair of agents determines the set of social connections and, consequently, the resulting network topology. Fig.~\ref{fig:constructGG}(b) shows a successful connection test between nodes $0$ and $6$. Fig.~\ref{fig:constructGG}(b) depicts a successful connection test between nodes 0 and 6. Since no other node lies inside the disk, illustrated by the hatched green circle, whose diameter is the segment joining the pair, they satisfy the Gabriel criterion, enabling the connection. Geometrically, this means that no intermediate agent spatially affects the direct interaction between the two nodes. Conversely, in panel (c), the candidate connection between nodes $0$ and $1$ is rejected because an intermediate node is present, as illustrated by the hatched red circle. Fig.~\ref{fig:constructGG}(d) shows the final Gabriel graph resulting from the successive application of the Gabriel criterion to every pair of nodes. The final network reflects a social planar proximity structure, with heterogeneous local neighborhoods and spatially constrained interactions.

To assemble connected Gabriel graphs of social networks, we initialize the system on a regular square lattice of linear size $L$, comprising $N = L^2$ nodes embedded in a two-dimensional unit square with periodic boundary conditions. The parameter $L$ specifies the number of lattice nodes along each direction, yielding a lattice spacing of $b = 1/L$, with one agent occupying each lattice site. To introduce spatial heterogeneity, we define a disorder parameter $s \in \mathbb{R}, \; 0 \leq s \leq 1$ that controls random displacements of agent positions. This network design displaces the coordinates $(x_i, y_i)$ of each agent $i$ as independent Gaussian random variables with zero mean and standard deviation $s \times b$, generated via the Box-Muller algorithm \cite{box1958note}. This formulation allows us to continuously interpolate the social interaction structure between regular square lattices ($s = 0$) and spatially disordered proximity networks generated from random agent placements ($s \neq 0$).

In Fig. \ref{fig:disorder_snet}, we illustrate five representative regimes of the spatial network organization generated by the disorder parameter $s$. In Fig. \ref{fig:disorder_snet}(a), corresponding to $ s = 0$, the Gabriel graph condition of Eq. (\ref{eq:gg}) recovers a regular square-lattice social network in which each agent interacts with exactly four nearest neighbors, producing a homogeneous and highly ordered interaction structure. In Fig. \ref{fig:disorder_snet}(b), for weak positional disorder, $s = 0.05$, small Gaussian perturbations modify the local geometry of the network, allowing additional diagonal connections and heterogeneous interaction neighborhoods to emerge while partially preserving the original lattice organization. As the disorder increases, shown in panel (c) for $s = 0.15$, the network enters a connectivity-enhanced regime characterized by increasingly heterogeneous local neighborhoods and larger average coordination. For stronger disorder levels, represented in panels (d) and (e), the positional randomness progressively suppresses the original geometric regularity. The interaction structure becomes increasingly irregular, producing heterogeneous spatial neighborhoods and random proximity connections. In this regime, interaction distances, neighborhood organization, and local connectivity become highly irregular, leading to a complex communication structure for opinion propagation.
 
\subsection*{Connectivity landscape and metrics}

We analyze the degree distribution, average connectivity, clustering coefficient, and the average path length of Gabriel graphs with variable disorder to understand how geometric heterogeneity shapes interaction patterns and, consequently, collective social organization. The degree distribution is the probability that an individual has a given number of social connections and provides a statistical description of the interaction landscape. The average connectivity quantifies the typical number of social contacts per individual and measures overall interaction density. The clustering coefficient quantifies the tendency of neighboring individuals to form locally interconnected groups, thereby promoting the reinforcement of information and the emergence of local consensus. In contrast, the average path length measures the typical separation between individuals in the network and therefore characterizes the efficiency of information propagation at the global scale.

The degree distribution $p(k)$ represents the probability that a randomly selected agent possesses $k$ social connections. As the disorder parameter $s$ increases, variations in $p(k)$ reveal how geometric heterogeneity redistributes social connections and modifies the local interaction environment experienced by individual agents,

\begin{equation}\label{eq:pk}
p(k)=\frac{N_k}{N},
\end{equation}
where $N_k$ is the number of agents with degree $k$

The average connectivity $\langle k \rangle$ quantifies the typical number of social contacts per individual,

\begin{equation}\label{eq:kmed}
\langle k \rangle = \frac{1}{N}\sum_{i=1}^{N} k_i,
\end{equation}
and larger values of $\langle k \rangle$ represent increased collective coordination and generally promote consensus formation \cite{2003CamposSmallWorld, 2005PereiraMajorityVote, 2006LuzMajorityVote}.

In the context of opinion dynamics, highly clustered regions may favor cohesive local communities by promoting opinion alignment among neighboring individuals. The local clustering of agent $i$ is 

\begin{equation}\label{eq:cci}
   \mathrm{C}_i = \frac{2e_i}{k_i(k_i-1)},
\end{equation}
in which $e_i$ is the number of edges that exist among the $k_i$ neighbors. To characterize the network structure, we use the average clustering coefficient

\begin{equation}\label{eq:ccMed}
\langle \mathrm{C} \rangle = \frac{1}{N}\sum_{i=1}^{N} \mathrm{C}_i.
\end{equation}
Variations in the disorder parameter $s$ alter the geometric arrangement of agents, affecting the formation of locally interconnected communities and the patterns of social interaction throughout the network.

The average path length is another essential network metric as it quantifies the typical number of interactions required for information to propagate between two arbitrary individuals in the social network. The average path length is

\begin{equation}\label{eq:apl}
\langle \ell \rangle= \frac{1}{N(N-1)} \sum_{i\neq j} \ell_{ij},
\end{equation}
where $\ell_{ij}$ is the minimal number of links connecting agent $i$ to agent\ $j$. In opinion dynamics, this quantity provides a direct measure of the efficiency of global communication and long-range social influence. Networks with short average path lengths facilitate faster information dissemination across distant regions of the system, promoting collective coordination and large-scale consensus formation. In contrast, larger path lengths hamper information flow, favoring fragmented social organization and slower convergence toward global consensus.

\subsection*{Opinion-updating rule}
Agents interact through a proximity-based communication network and update their opinions by observing the states of their local neighbors. Social influence operates through a tendency toward conformity, whereby individuals tend to align with the majority opinion in their neighborhood while retaining a finite probability of deviation. This stochastic behavior is captured by the majority-vote model, in which each agent $i$ adopts the local majority opinion with probability $1 - q$ and, with probability $q$, adopts the opposite state, representing independent or contrarian behavior. In the absence of a local majority, when the neighborhood contains an equal number of opposing opinions, agents choose one of the two states $\sigma_i = \pm 1$ with equal probability \cite{1992Oliveira_Isotropic}.

We denote the social influence group of agent $i$ by $\Lambda_i$, which consists of all individuals directly connected to it in the network. The spin-flip probability, which describes the stochastic dynamics of the opinion variable $\sigma_i$ is

\begin{equation}\label{eq:mvm}
w_i(\sigma) = \frac{1}{2} \left[1 - (1 - 2q)\,\sigma_i\,\operatorname{sgn}\!\left(\sum_{j\in\Lambda_i} \sigma_j \right) \right],
\end{equation}
where $\text{sgn}(a) = -1, 0, +1$ for $a < 0$, $a = 0$, and $a > 0$, respectively. The summation runs over all agents connected to agent $i$ through the Gabriel graph interaction network. This transition rate defines a nonequilibrium dynamics controlled by a nonconformity parameter $q$, with $0 \leq q \leq 1/2$.

\subsection*{Quantifying emergent collective behavior}

We assess the emergence of collective social order by estimating the instantaneous absolute mean opinion of the population as the order parameter of the system, analogous to a magnetization of a spin model

\begin{equation}\label{eq:orderparameter}
    m(t) = \frac{1}{N} \left| \sum_{i=1}^{N} \sigma_i(t)\right|.
\end{equation}
This quantity allows us to measure the degree of collective alignment among individuals. Values of $m \approx 1$ denote highly ordered regimes in which most individuals share the same opinion, reflecting the emergence of a macroscopic social consensus. In contrast, values of $m \approx 0$ indicate that opposite opinions are statistically balanced, corresponding to a socially disordered state without a dominant collective orientation.

Due to the stochastic nature of both the opinion dynamics and the spatial network construction, we average physical observables over time and across independent network realizations to obtain statistically representative macroscopic behavior. We define below the primary quantities that characterize the system's collective state. The average magnetization is

\begin{equation}\label{eq:MSU}
    M(q,N,s) = \left\langle \left\langle m \right\rangle_t \right\rangle_c,
\end{equation}
and quantifies the average level of global consensus in the system as a function of the nonconformity parameter $q$, the system size $N$, and the positional disorder parameter $s$. Large values of $M$ indicate strongly ordered social states dominated by a majority opinion, whereas small values signal fragmented or socially disordered configurations. The notation $\langle \cdot \rangle_t$ indicates averages over time in the stationary regime, whereas $\langle \cdot \rangle_c$ denotes averages over independent network realizations.

We also measure fluctuations in the collective opinion to estimate the consensus sensitivity to stochastic perturbations using the opinion susceptibility
\begin{equation}
    \chi(q,N,s) = N \left[
    \left\langle \left\langle m^2 \right\rangle_t \right\rangle_c
    - \left\langle \left\langle m \right\rangle_t \right\rangle_c^2
    \right].
\end{equation}
Near the transition between consensus and dissent, small local opinion changes can trigger large collective responses, producing pronounced peaks in $\chi$. Consequently, the susceptibility serves as an important indicator of critical behavior driven by variations in the nonconformity parameter $q$ and the spatial disorder parameter $s$.

In general, collective phenomena depend on the number of interacting agents, and finite-size effects directly influence the characterization of the system near the thermodynamic limit. To accurately identify the critical point and investigate finite-size scaling properties, we compute Binder’s fourth-order cumulant

\begin{equation}
    U(q,N,s) = 1 -
    \frac{
    \left\langle \left\langle m^4 \right\rangle_t \right\rangle_c
    }{
    3\left\langle \left\langle m^2 \right\rangle_t \right\rangle_c^2
    },
\end{equation}
For different system sizes $N$, the cumulant curves intersect near the critical nonconformity parameter $q_c$, allowing a robust estimation of the phase-transition point and its universality properties.

\subsection*{Simulation details}
We perform extensive Monte Carlo simulations across multiple system sizes, nonconformity levels, and disorder regimes. We generate on spatially embedded networks from Gaussian perturbations of an initial square lattice configuration, with the disorder parameter $s$ continuously controlling the degree of positional randomness. For each realization, the opinion system evolves asynchronously through stochastic opinion updates governed by the nonconformity parameter $q$, where we ignore the transient relaxation interval $\tau_{\mathrm{relax}}$, and perform measurements in the stationary regime over a measurement interval $\tau_{\mathrm{meas}}$. To reduce autocorrelation effects between successive measurements, we sample quantities at every $\tau_{\mathrm{skip}}$ Monte Carlo steps. We compute statistical averages over $\eta_{\mathrm{nets}}$ independent network realizations, with $\eta_{\mathrm{runs}}$ simulation runs per network configuration.

Table~\ref{tab:simParameters} summarizes the system sizes, disorder and nonconformity ranges, Monte Carlo timescales, and averaging procedures of the majority-vote model on disordered Gabriel graph social networks.

\begin{table}[!ht]
\centering
\caption{Simulation parameters and numerical setup}
\label{tab:simParameters}
\begin{tabular}{lll}
\hline
\textbf{Name} & \textbf{Description} & \textbf{Value} \\
\hline

$L$ & Linear size & $40$ to $120$ \\

$N$ & Number of agents & $1600$ to $14400$ \\

$A$ & Cartesian area & $1.0 \times 1.0$ \\

$s$ & Disorder parameter range & $[0.0, 1.0]$ \\

$q$ & Nonconformity parameter range & $[0.0, 0.5]$ \\

$\tau_{\mathrm{relax}}$ & Relaxation time & $3 \times 10^5$ \\

$\tau_\mathrm{{meas}}$ & Measurement time & $10^5$ \\

$\tau_\mathrm{{skip}}$ & Decorrelation interval & $10^2$ \\

$\eta_{\mathrm{nets}}$ & Number of networks & $10^4$ \\

$\eta_{\mathrm{runs}}$ & Number of runs & $10^2$ \\

\hline
\end{tabular}
\end{table}

\section*{Results}
\label{sec:results}

In this section, we investigate structural regimes of Gabriel graph social networks with spatial disorder by analyzing the degree distribution, clustering coefficient, and average path length. These quantities allow us to characterize how geometric heterogeneity reshapes the interaction network and influences local social organization and consensus formation. We then examine how spatial disorder, geometric constraints, social nonconformity, and system size collectively affect the opinion dynamics. In particular, we identify consensus-dissent phase transitions emerging from spatial disorder and nonconformity levels. Finally, we characterize critical behaviors by estimating critical exponents using volumetric and linear finite-size scaling techniques.

\subsection*{Connectivity transition on disordered social proximity networks}

\begin{figure*}[!ht]
	\centering
	\includegraphics[width=1.0\linewidth]{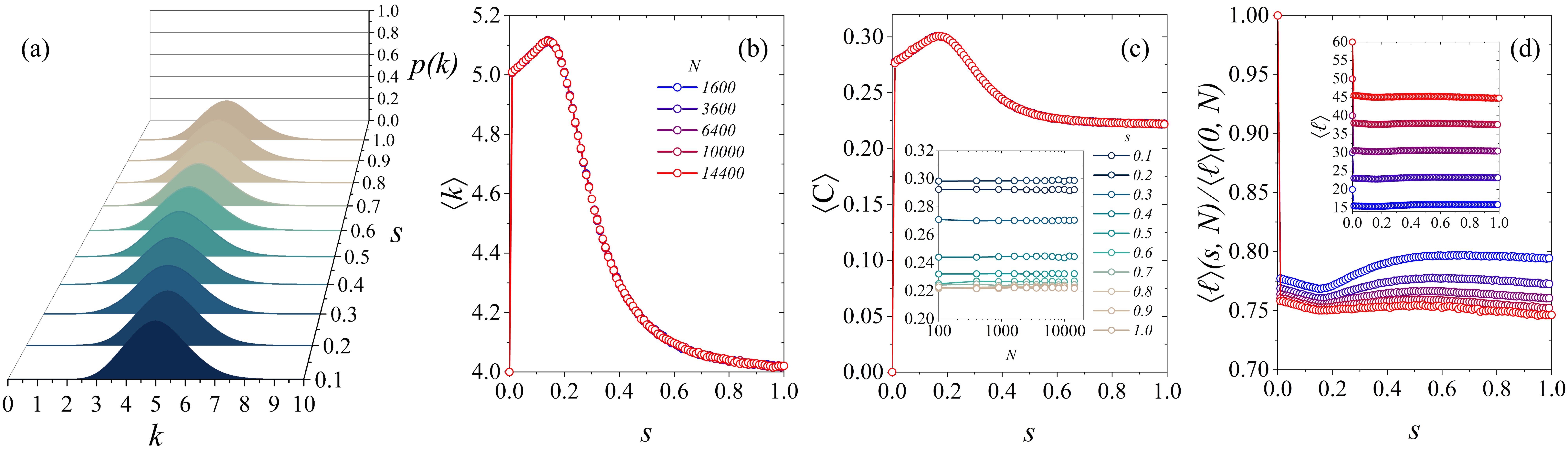}
	\caption{\textbf{Structural properties of Gabriel graph social networks under spatial disorder.} (a) Dependence of the degree distribution $p(k)$ with increasing disorder parameter $s$ for $N = 14400$. Weak disorder shifts the distributions toward larger connectivities, while stronger disorder progressively broadens the distributions and moves them toward lower coordination values. (b) Average connectivity $\langle k \rangle$ as a function of $s$ for different system sizes $N$, revealing a disorder-induced connectivity enhancement followed by a gradual suppression toward the spatially random regime. (c) Average clustering coefficient $\langle C \rangle$ as a function of $s$. Weak geometric disorder initially increases local clustering, while stronger disorder stabilizes the network into heterogeneous proximity-based communities. The inset shows the weak dependence of $\langle C \rangle$ on system size. (d) Average shortest-path length normalized by system size, $\langle \ell \rangle (s, N)/ \ell(0, N)$, as a function of $s$. Weak positional disorder initially reduces path distances due to the emergence of additional local shortcuts, whereas stronger disorder preserves relatively large path lengths. The inset displays the corresponding unnormalized average path lengths $\langle \ell \rangle$. The results are averaged over up to $\eta_{\mathrm{nets}} = 10^4$ independent network realizations. The lines are guides to the eye.}
	\label{fig:pkkmed}
\end{figure*}

The disorder parameter $s$ induces critical structural changes that directly affect the reinforcement of local opinion and the efficiency of long-range communication. In Fig.~\ref{fig:pkkmed}, we illustrate how introducing positional disorder profoundly reshapes the structural organization of Gabriel graph social networks. Fig.~\ref{fig:pkkmed}(a) shows degree distributions $p(k)$ that reveal the variations of connectivity structure as disorder increases. For weak disorder levels, $s \gtrsim 0$, geometric perturbations introduce diagonal connections and heterogeneous neighborhoods, broadening the degree distribution and shifting its maximum toward larger connectivities. This regime corresponds to a disorder-induced enhancement of connectivity, in which small positional perturbations increase the network's average coordination. As disorder increases further, $s \approx 1$, the degree distribution progressively broadens while the average connectivity decreases toward the random Gabriel graph regime.

The average connectivity shown in Fig.~\ref{fig:pkkmed}(b) highlights a nonmonotonic response to positional disorder and confirms the results of panel (a). Starting from the square-lattice case with $s = 0$ and $k = 4$ for all agents, weak disorder initially increases the average connectivity, reaching a maximum near $s = 0.15$. In this regime, local geometric perturbations create additional diagonal short-range links, yielding $\langle k \rangle \approx 5.1$. Beyond this point, increasing disorder progressively suppresses average connectivity as the network transitions toward a spatially randomized proximity structure. We highlight that all system sizes collapse onto a single curve, indicating that the connectivity transitions are robust to finite-size effects and are primarily governed by the disorder parameter and geometric constraints imposed by the Gabriel condition.

The connectivity transitions driven by $s$ deeply influence the mechanisms of information propagation and collective social organization. The regular lattice regime favors homogeneous local interactions due to the absence of structural heterogeneity. In contrast, weak positional disorder enhances local connectivity and creates alternative communication pathways, facilitating both local consensus formation and more efficient opinion propagation across the network. At high disorder levels, the system forms highly heterogeneous spatial neighborhoods while preserving the proximity constraints. As a consequence, the network combines localized community organization with geometrically constrained communication, creating a complex interaction structure for the emergence of collective opinion dynamics.

\subsection*{Local cohesion and information propagation}

Fig.~\ref{fig:pkkmed}(c) shows the average clustering coefficient $\langle C \rangle$, which quantifies the tendency of neighboring agents to form locally interconnected groups. In the regular lattice regime of $s = 0$, the clustering coefficient vanishes because square lattices do not contain triangular neighborhoods. Introducing weak disorder rapidly increases $\langle C \rangle$, reflecting the emergence of local triangular connections and community-like structures generated by positional perturbations. For larger disorder levels, the clustering coefficient stabilizes around an intermediate value, indicating that local social cohesion persists even in strongly disordered proximity networks. The inset further shows that $\langle C \rangle$ depends only weakly on the system size.

Finally, Fig.~\ref{fig:pkkmed}(d) presents the average shortest-path length normalized by its square-lattice value, $\langle \ell \rangle (s, N) / \ell(0, N)$. Weak disorder initially reduces characteristic path lengths because newly formed diagonal and irregular local connections create more efficient routes across the network. However, even at high disorder levels, the path length remains relatively large compared with that of small-world or random networks \cite{2016barabasi_network}, demonstrating that spatial constraints continue to limit the efficiency of long-range communication. The inset showing the unnormalized path lengths confirms that information transport remains strongly influenced by the network's geometric embedding, even in highly disordered regimes. Overall, this transitional regime of Gabriel graph social networks enables a flexible transmission via local clustering and heterogeneity; however, the absence of long-range shortcuts still limits global diffusion.

\subsection*{Collective ordering under social nonconformity}

\begin{figure}[!ht]
	\centering
	\includegraphics[width=1.00\linewidth]{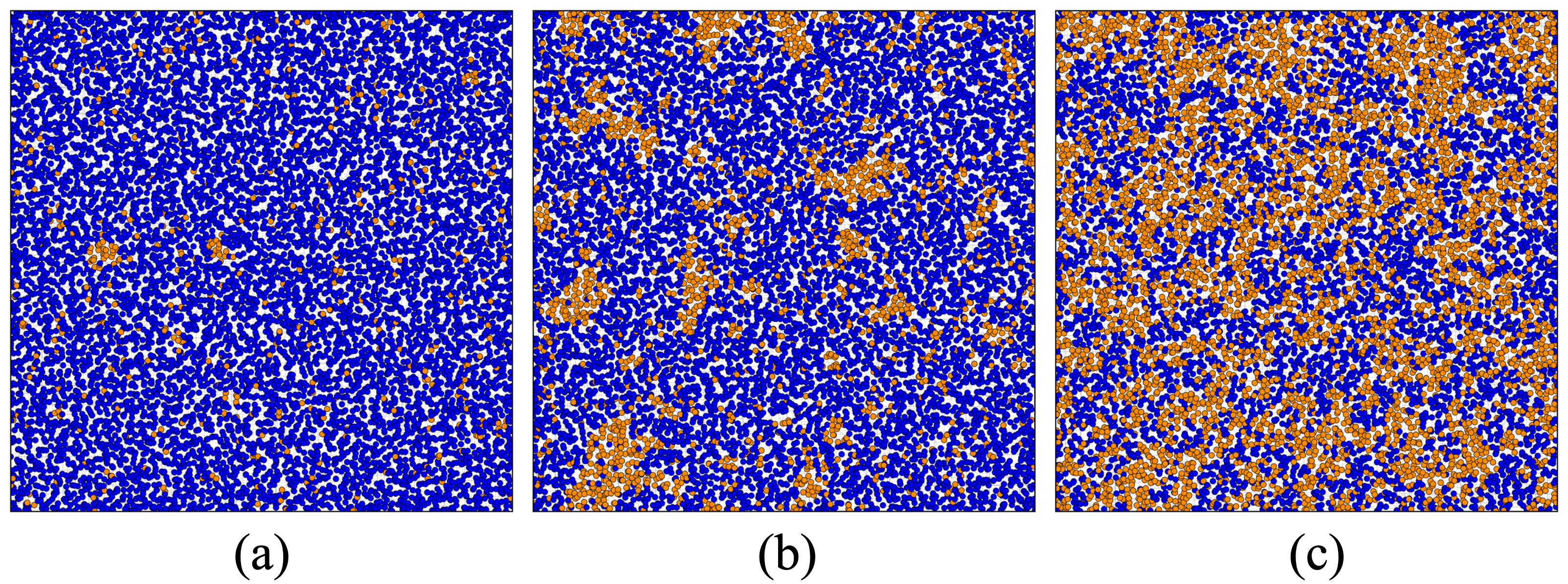}
	\caption{\textbf{Collective opinion patterns for increasing nonconformity levels.} Snapshots of a simulation on a Gabriel graph network with $s = 0.5$, $N = 10^4$ and increasing nonconformity $q$. Blue and orange nodes represent agents holding opposite opinions. (a) Ordered consensus regime at $q = 0.036$. (b) Critical region near the consensus--dissent transition at $q = 0.073$. (c) Disordered dissensus regime at $q = 0.146$.}
	\label{fig:snapshots}
\end{figure}

Large-scale opinion organization emerges from the competition between social pressure and stochastic disagreement. In the majority-vote model, this dynamics gives rise to a nonequilibrium consensus--dissent transition at a critical nonconformity level $q_c$, separating an ordered phase characterized by macroscopic collective alignment from a disordered regime dominated by competing coexisting opinions. To characterize the emergence and persistence of collective order, we investigate the dependence of the magnetization $M(q,N,s)$, susceptibility $\chi(q,N,s)$, and Binder cumulant $U(q,N,s)$ on the social nonconformity parameter $q$ and geometric disorder parameter $s$.

The Fig. \ref{fig:snapshots} illustrates the evolution of collective opinion organization as the nonconformity level increases for a geometric disorder level $s = 0.5$ and $N = 10^4$. The snapshots reveal how increasing nonconformity progressively destroys global consensus while preserving strong spatial correlations induced by proximity-based interactions. At low levels of nonconformity, local majority interactions dominate the dynamics, leading to a globally ordered consensus state in which one opinion prevails throughout the system. As $q$ increases, stochastic disagreement becomes more relevant, and domains of opposite opinions progressively arise, signaling the onset of critical fluctuations near the consensus--dissent transition. For sufficiently high $q$, the system loses long-range order and evolves into a fragmented dissensus regime characterized by competing spatial opinion clusters and the absence of a dominant collective orientation. 

\begin{figure*}[!ht]
    \centering \includegraphics[width=1.0\linewidth]{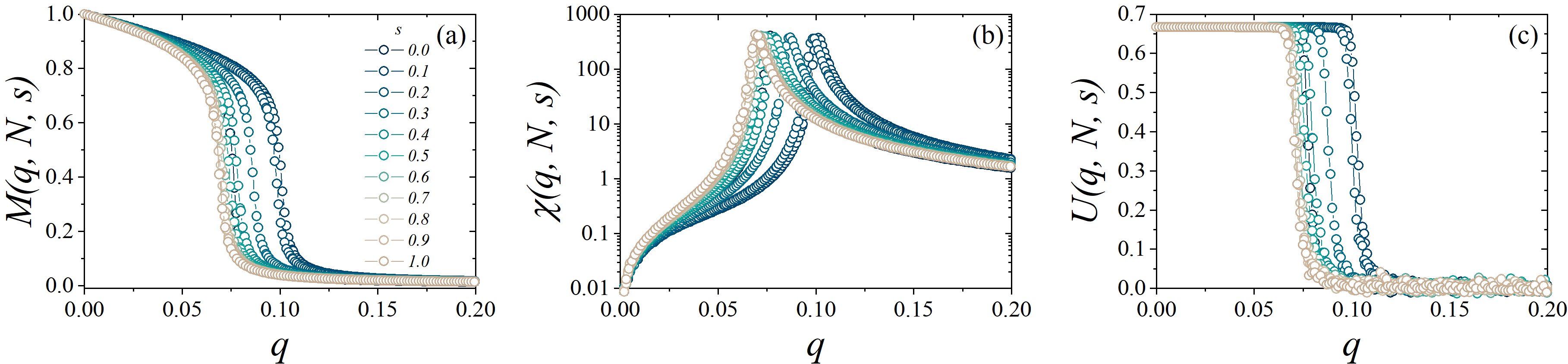}
    \caption{\textbf{Effects of spatial disorder on collective opinion dynamics.} (a) Magnetization $M(q, N, s)$, (b) opinion susceptibility $\chi(q, N, s)$, and (c) Binder’s fourth-order cumulant $U(q, N, s)$ as functions of the nonconformity parameter $q$ for different disorder levels $s$ and $N = 14400$. Increasing spatial disorder shifts the location of the consensus--dissent transition, initially toward larger values of $q$ for weak disorder, and progressively back toward lower values for strong disorder. The susceptibility peaks allow for an estimation of critical nonconformity levels, while the Binder cumulant provides a robust characterization of the transition region. The lines are guides to the eye.}
    \label{fig:mxuvss}
\end{figure*}

In Fig.~\ref{fig:mxuvss}, we show the effects of spatial disorder on magnetization $M(q, N, s)$, opinion susceptibility $\chi(q, N, s)$, and (c) Binder’s fourth-order cumulant $U(q, N, s)$ as functions of the nonconformity parameter $q$ for different disorder levels $s$ and $N = 14400$. All curves in Fig.~\ref{fig:mxuvss}(a) exhibit the same qualitative behavior, revealing a transition from complete consensus to social disorder as the magnetization $M$ decays from $1$ to approximately $0$ with increasing $q$. Regardless of the disorder level, the case $q = 0$ corresponds to a fully ordered social state in which all individuals share the same opinion and $M = 1$. As nonconformity increases, stochastic deviations from the local majority progressively fragment opinions across the population, yielding $M \to 0$ and weakening global consensus.

A notable effect of spatial disorder is the displacement of the transition region along the nonconformity axis. Weak geometric disorder initially shifts the transition toward larger $q$ values, indicating that moderate positional heterogeneity enhances the robustness of consensus against stochastic disagreement. For stronger disorder levels, however, the transition progressively shifts back toward lower values of $q$, suggesting that highly disordered interaction networks reduce the robustness of collective social organization. At sufficiently large nonconformity, competing opinions coexist macroscopically within the population, characterizing a socially disordered regime. The continuous decay of the magnetization and the absence of abrupt discontinuities indicate second-order nonequilibrium phase transitions between ordered and disordered social states.

The susceptibility curves of Fig.~\ref{fig:mxuvss}(b) reveal that spatial disorder significantly affects the location of the transition region. The low-susceptibility regions correspond to social states of strong consensus at low $q$ and of expressive disagreement at large $q$. In contrast, susceptibility peaks identify the transition region dominated by local competing social influences. As in the magnetization, weak geometric disorder shifts the susceptibility peak toward higher $q$ values, indicating that moderate positional heterogeneity increases the robustness of consensus against stochastic disagreement. For higher disorder levels, however, the peaks shift back toward lower values of $q$, suggesting that highly disordered interaction networks reduce the stability of collective social organization.

Fig.~\ref{fig:mxuvss}(c) shows Binder’s fourth-order cumulant $U(q, N, s)$ that exhibits an abrupt variation near the transition region, separating the ordered consensus phase from the socially disordered regime. At low levels of nonconformity, the cumulant remains close to its maximum, reflecting the dominance of a globally ordered social state with strong collective alignment. As the nonconformity parameter increases, the cumulant rapidly decreases toward values near zero, indicating the suppression of global consensus and the emergence of competing coexisting opinions.

\subsection*{System-size effects on consensus and dissent}

Complex collective phenomena naturally depend on the number of interacting agents, and finite-size effects play an important role in the emergence of social consensus and critical behavior. In general, finite populations display phase transitions that do not occur as perfectly sharp discontinuities but rather as smooth transitions whose signatures become progressively more pronounced as the system size increases. Consequently, quantities such as the magnetization and susceptibility exhibit systematic size-dependent behavior near the transition region, allowing the characterization of collective ordering and the extrapolation of the system behavior toward the thermodynamic limit. To investigate these effects, we analyze the dependence of magnetization $M(q, N, s)$, opinion susceptibility $\chi(q, N, s)$, and Binder cumulant $U(q, N, s)$ on the system size $N$.

\begin{figure*}[!ht]
    \centering \includegraphics[width=1.0\linewidth]{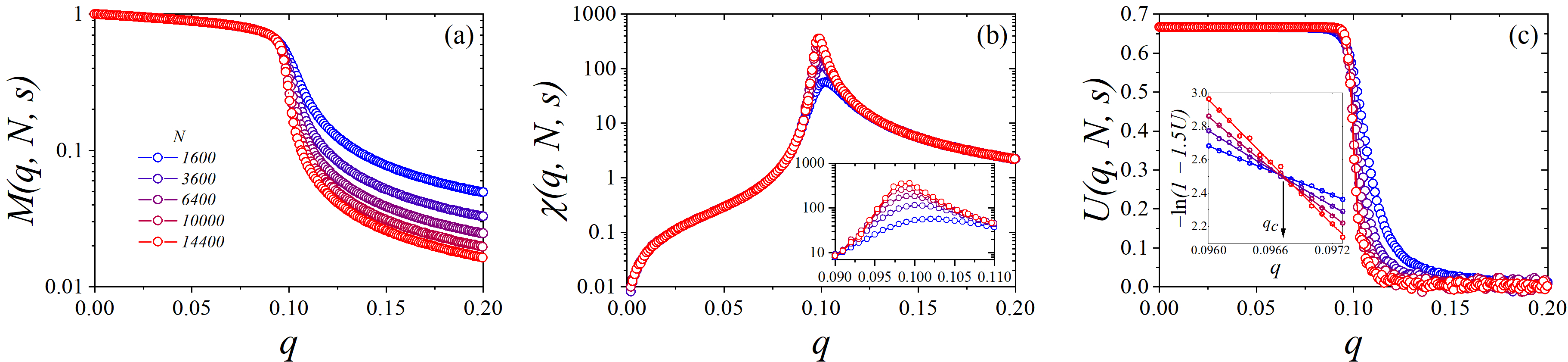}
    \caption{\textbf{Finite-size effects on collective opinion dynamics.}
(a) Magnetization $M(q,N,s)$, (b) opinion susceptibility $\chi(q,N,s)$, and (c) Binder’s fourth-order cumulant $U(q,N,s)$ as functions of the nonconformity parameter $q$ for different system sizes and social disorder level $s = 0.2$. Increasing system size sharpens the transition from consensus to dissent and enhances collective fluctuations near criticality. The susceptibility peaks provide estimates for critical nonconformity levels $q_c(N,s)$. In contrast, the intersection of Binder cumulant curves provides a robust estimation of the critical point at the thermodynamic limit and confirms the second-order nonequilibrium nature of the phase transition. The inset in panel (b) highlights the growth of susceptibility peaks with increasing system size. The inset in panel (c) shows rescaled Binder cumulant $-\ln(1-1.5U)$ for different system sizes. The lines are guides to the eye.}
    \label{fig:mxuvsN}
\end{figure*}

In Fig. \ref{fig:mxuvsN}(a), the magnetization curves for $s = 0.2$ exhibit a continuous decay from complete consensus toward social disorder as the nonconformity parameter $q$ increases. The transition region becomes progressively steeper for larger systems, indicating that large societies exhibit increasingly sharp changes in collective opinion organization. Simultaneously, the residual magnetization observed beyond the transition region decreases with system size, showing that the socially disordered regime becomes more evident as the system approaches the thermodynamic limit $N \to \infty$.

The susceptibility curves of Fig. \ref{fig:mxuvsN}(b) further highlight the emergence of size-dependent critical collective behavior for a disorder level $s = 0.2$. Near the transition region, the susceptibility exhibits a pronounced maximum, reflecting the amplification of collective fluctuations and the population's increasing sensitivity to local opinion perturbations. As the system size increases, these peaks become significantly sharper and higher, revealing the strengthening of collective critical behavior in large populations. The systematic dependence of magnetization and susceptibility on system size provides important information about critical behavior, clearly indicating the existence of distinct macroscopic phases.

 We leverage finite-size effects of Binder's fourth-order cumulant to estimate the critical nonconformity level in the thermodynamic limit. Since the cumulant exhibits weak size dependence near the critical region, curves corresponding to different system sizes intersect approximately at a common point, allowing a robust estimation of the critical nonconformity parameter $q_c$, separating the ordered consensus regime from the socially disordered phase. 
 
Fig.~\ref{fig:mxuvsN}(c) presents the Binder fourth-order cumulant for different system sizes and $s = 0.2$. The sharp drop of the cumulant marks the breakdown of macroscopic consensus and the onset of the socially disordered phase. Below the critical nonconformity level, the cumulant remains approximately constant, reflecting the robustness of the ordered collective state. As the nonconformity parameter approaches the critical region, the cumulant rapidly decreases due to the substantial growth in collective fluctuations and the loss of global opinion alignment. The inset highlights the intersection of the transformed cumulant data with cubic interpolation fits across different system sizes, enabling an accurate estimate of the critical nonconformity that suppresses global consensus in the thermodynamic limit. For $s = 0.2$, we obtain $q_c \approx 0.0967(4)$.

\subsection*{Collective phases in disordered social proximity networks}

\begin{figure}[!hb]
	\centering
	\includegraphics[width=0.9\linewidth]{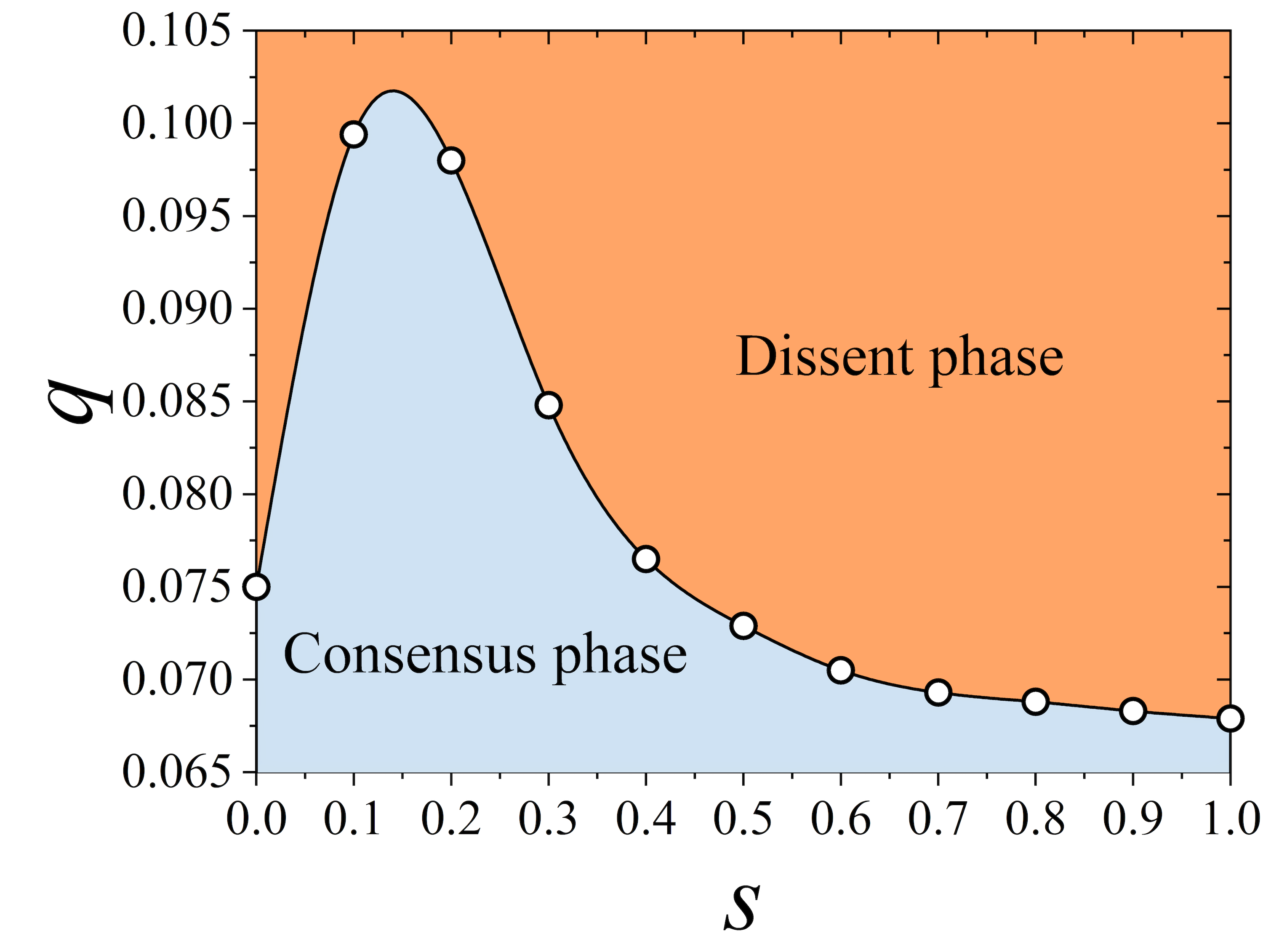}
	\caption{\textbf{consensus--dissent phase diagram under spatial disorder.} Critical nonconformity values as a function of the disorder parameter for the majority-vote dynamics on Gabriel graph social networks. The phase boundary separates the ordered phase, characterized by macroscopic social consensus, from the disordered phase, where competing opinions coexist without global alignment. Weak geometric disorder initially increases the critical nonconformity, indicating enhanced consensus robustness due to the connectivity-enhanced regime, where increased local coordination and clustering strengthen collective social organization. For higher disorder levels, however, the critical threshold decreases progressively as the network approaches the spatially random regime, reflecting the reduced stability of collective social organization under strong geometric heterogeneity. The error bars are smaller than the symbol size.}
	\label{fig:phaseDiagram}
\end{figure}

Phase diagrams provide a comprehensive representation of the collective states accessible to a system as control parameters change. In nonequilibrium opinion dynamics, they allow us to identify how microscopic interaction mechanisms and network structure determine the emergence, stability, and decline of macroscopic social organization. Fig.~\ref{fig:phaseDiagram} presents the nonequilibrium phase diagram of the majority-vote dynamics on Gabriel graph social networks as spatial disorder increases. Binder cumulant calculations provide estimations of the critical boundary $q = q_c(s)$ that separates the ordered consensus phase from the socially disordered regime. Below the critical boundary, the system remains in an ordered phase characterized by the macroscopic predominance of a collective opinion. In this regime, local majority interactions overcome stochastic disagreement, enabling global social consensus to emerge and persist across the population. Above the transition line, however, social nonconformity becomes sufficiently strong to destabilize the collective organization, producing a disordered phase in which competing opinions coexist macroscopically without long-range alignment. 

The phase boundary separates regions characterized by qualitatively distinct collective behaviors and reveals the effects of structural heterogeneity: moderate disorder promotes consensus, whereas strong disorder destabilizes global social organization. A notable aspect of the phase diagram is the nonmonotonic dependence of the critical nonconformity on the disorder parameter $s$. Slight geometric disorder levels initially shift the transition to larger values of $q_c$, reaching a maximum near $s \approx 0.15$. This behavior indicates that moderate positional heterogeneity enhances the robustness of collective consensus against stochastic disagreement. From a structural perspective, this regime resembles the connectivity-enhanced phase of Gabriel graph networks shown in Fig. \ref{fig:pkkmed}(b), in which increased local coordination and clustering strengthen collective social organization.

Nevertheless, for higher disorder levels, the critical threshold decreases progressively as the network approaches the intensely disordered spatial-random regime. In this region, excessive geometric heterogeneity weakens the efficiency of collective social organization by suppressing the enhanced connectivity observed at intermediate levels of disorder and reducing the effectiveness of local support mechanisms. Consequently, consensus becomes increasingly fragile as social nonconformity grows.

\subsection*{Finite-size scaling and universality}
Finite-size effects greatly impact collective observables near continuous phase transitions. Close to criticality, systems belonging to the same universality class exhibit scale-invariant behavior characterized by universal scaling functions and critical exponents, independent of microscopic details \cite{Stanley1987PhaseTransitions}. This property enables the characterization of the consensus--dissent transition via finite-size scaling theory, which connects the behavior of finite systems to the critical behavior expected in infinitely large populations.

For regular lattices of dimension $d$, the number of nodes contained within a region of characteristic size $\xi$ scales as $n \sim \xi^{d}$. Near the critical point, the correlation length becomes comparable to the system size, implying $n \sim N$, where $N$ is the total number of nodes. This observation naturally motivates using $N$ as a scaling variable \cite{2020VilelaThree}. Hence, the critical behavior of the magnetization, susceptibility, and Binder cumulant obeys the volumetric scaling relations

\begin{equation}\label{eq:scalingM}
M(q,N,s)=N^{-\beta/\bar{\nu}}\widetilde{M}(\epsilon N^{1/\bar{\nu}}),
\end{equation}

\begin{equation}\label{eq:scalingX}
\chi(q,N,s) = N^{\gamma/\bar{\nu}}\widetilde{\chi}(\epsilon N^{1/\bar{\nu}}),
\end{equation}

\begin{equation}\label{eq:scalingU}
U(q,N,s) = \widetilde{U}(\epsilon N^{1/\bar{\nu}}),
\end{equation}
where $\epsilon = q - q_c$ measures the deviation from the critical nonconformity level. The functions $\widetilde{M}$, $\widetilde{\chi}$, and $\widetilde{U}$ are universal scaling functions associated with the critical behavior. The exponent ratios $1/\bar{\nu}$, $\beta/\bar{\nu}$, and $\gamma/\bar{\nu}$ characterize the universality class of the system and determine how collective fluctuations scale with the system size $N$.

Within the volumetric scaling framework, the critical exponents satisfy the unitary relation regardless of the network structure \cite{2020VilelaThree},

\begin{equation}\label{eq:unit}
2\frac{\beta}{\bar{\nu}} + \frac{\gamma}{\bar{\nu}} \equiv \upsilon = 1,
\end{equation}
which provides an important consistency condition for the scaling analysis. This relation reflects the normalization of collective fluctuations with respect to the system volume and is particularly appropriate for systems with complex or heterogeneous interaction topologies.

For completeness, we also investigate the standard hyperscaling relation by analyzing the critical behavior of Eqs. (\ref{eq:scalingM}), (\ref{eq:scalingX}), (\ref{eq:scalingU}) as functions of the linear size $L=\sqrt{N}$, using the standard critical exponents $\beta$, $\gamma$, and $\nu$ \cite{Stanley1987PhaseTransitions}

\begin{equation}\label{eq:hype}
2\frac{\beta}{\nu} + \frac{\gamma}{\nu} = d.
\end{equation}
This analysis allows us to verify whether the nonequilibrium critical behavior of the majority-vote dynamics on Gabriel graph social networks remains compatible with the hyperscaling properties of equilibrium critical systems.

\begin{figure*}[!ht]
	\centering
	\includegraphics[width=1.0\linewidth]{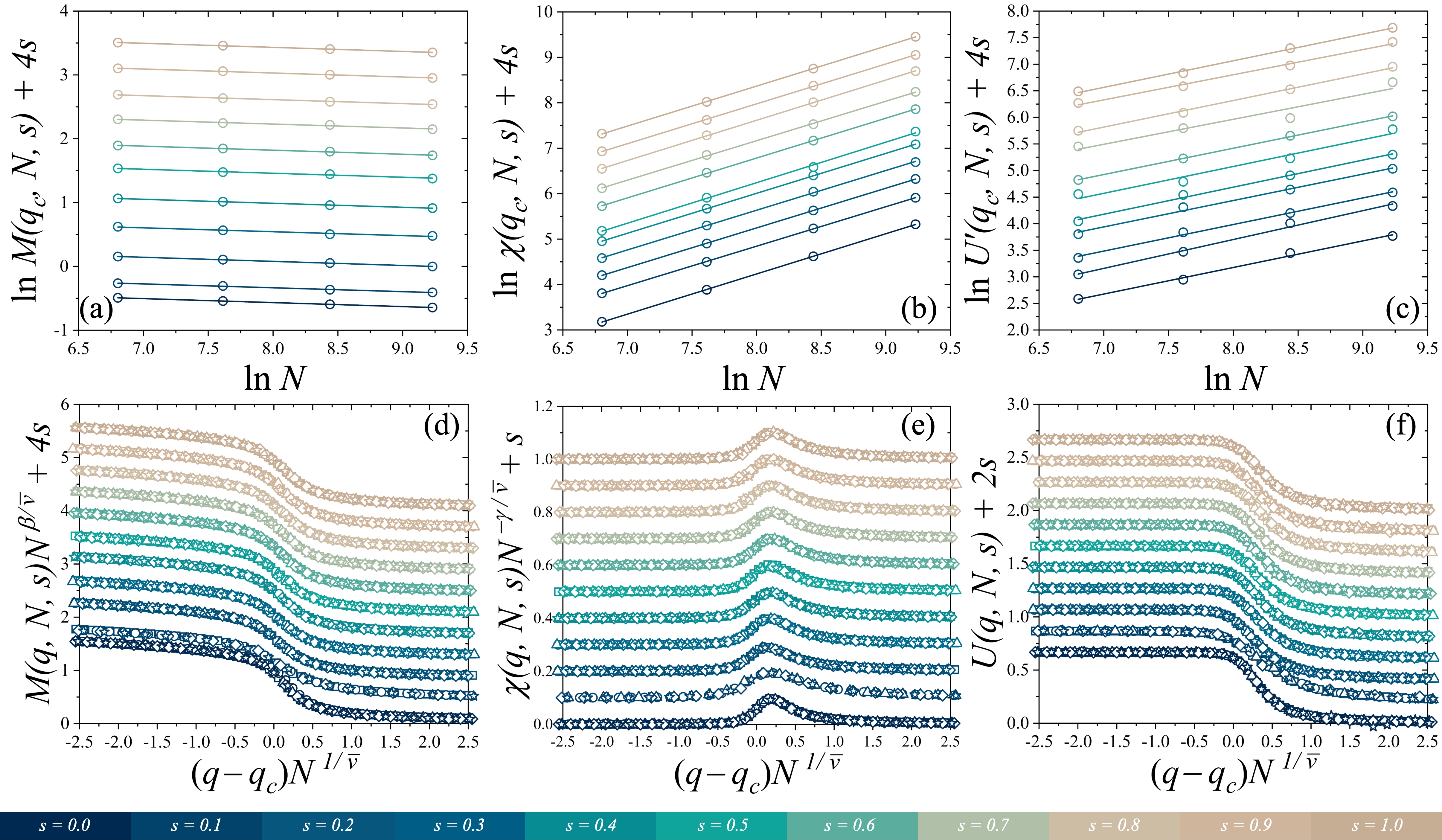}
	\caption{\textbf{Finite-size scaling and data collapse in volumetric representation.} Panels (a), (b), and (c) show log-log plot at the critical social temperature $q_c(s)$ as a function of the system size $N$: (a) average opinion $M(q_c,N,s)$, (b) susceptibility $\chi(q_c,N,s)$, and (c) derivative of Binder’s cumulant $U'(q_c,N,s)$. Panels (d), (e), and (f) present the data collapse obtained using volumetric scaling, where the control parameter is rescaled as $(q-q_c)N^{1/\bar{\nu}}$, for (d) average opinion, (e) susceptibility, and (f) Binder's cumulant, in which each curve (color) comprises the network sizes $N = 1600, 3600, 6400, 10000$, and $14400$. We shift each set of data symbols to avoid overlaps for different values of $s$ for clarity. The error bars are smaller than the symbol size.
	}
	\label{fig:dataCollapse-expoN}
\end{figure*}

\subsection*{Critical exponents and data collapse}

Critical exponents are crucial for classifying universality classes and for identifying whether different interaction structures preserve or alter underlying critical properties of the system. To estimate the critical exponents, we analyze the finite-size behavior of the collective observables precisely at the critical point, where scale invariance emerges, and the finite-size scaling relations reduce to power laws. At $q = q_c$, the deviation from criticality vanishes ($\epsilon=0$), eliminating the dependence on the scaling functions and allowing the critical exponents to be directly extracted from the system-size dependence of the observables. Taking the logarithm of the volumetric finite-size scaling relations yields linear behaviors of the form

\begin{equation}\label{eq:scalinglnM}
\ln [M(q_c, N, s)] \sim -\frac{\beta}{\bar{\nu}} \ln [N]
\end{equation}

\begin{equation}\label{eq:scalinglnX}
\ln [\chi(q_c, N, s)] \sim \frac{\gamma}{\bar{\nu}} \ln [N],
\end{equation}

\begin{equation}\label{eq:scalinglnU}
\ln [U(q_c, N, s)] \sim \frac{1}{\bar \nu} \ln [N],
\end{equation}
from which the exponent ratios are estimated through the slopes of the corresponding log-log curves. This procedure allows the characterization of the universality properties of the consensus–dissent transition and provides a direct verification of the unitary relation of Eq. (\ref{eq:unit}).

Fig. \ref{fig:dataCollapse-expoN} presents the finite-size scaling analysis and volumetric data collapse for the majority-vote dynamics on disordered Gabriel graph social networks. The results demonstrate that the collective opinion transition satisfies universal scaling relations across all disorder regimes and provide strong evidence for the robustness of the volumetric scaling hypothesis. Panels (a), (b), and (c) show the log-log scaling behavior of the critical observables evaluated at the critical nonconformity $q_c(s)$. We displace each set of data symbols to avoid overlaps for different values of $s$. In Fig. \ref{fig:dataCollapse-expoN}(a), the average opinion $M(q_c, N, s)$ exhibits approximately linear behavior as a function of $\ln N$, consistent with the power-law scaling predicted by finite-size scaling theory. The negative slopes indicate the suppression of macroscopic consensus as the system size increases at criticality, allowing the estimation of the exponent ratio $\beta/\bar{\nu}$. Fig. \ref{fig:dataCollapse-expoN}(b) presents the susceptibility $\chi(q_c, N, s)$, whose linear growth with $\ln N$ reflects the amplification of collective opinion fluctuations near the transition region. The positive slopes provide estimates of the ratio $\gamma/\bar{\nu}$, which characterizes how critical fluctuations diverge with the system volume. In Fig. \ref{fig:dataCollapse-expoN}(c), the derivative of Binder’s cumulant also follows linear scaling behavior, enabling the extraction of $1/\bar{\nu}$, which governs the scaling of the correlation volume near criticality. Table \ref{tab:criticalExpoforDiffs_volumetric} summarizes the estimated critical nonconformity levels and critical exponent ratios for each disorder configuration. Despite the displacement of the transition point, the critical exponents remain compatible with the same \textit{volumetric} universality class throughout the entire disorder range.

\begin{table*}[!ht]
    \centering
    \begin{tabular}{|c|c|c|c|c|c|c|c|c|}\hline
    \rowcolor{myblue!100}
        \textcolor{white}{$s$} & \textcolor{white}{$q_c$} & \textcolor{white}{$\beta / \bar{\nu}$} & \textcolor{white}{$\gamma / \bar{\nu}$} & \textcolor{white}{$1 / \bar{\nu}$} & \textcolor{white}{$\upsilon$} \\ \hline\hline
        $0.0$ & $0.0750(1)$ & $0.0625(1)$ & $0.875(1)$ & $0.50(1)$ & $1.000(2)$ \\ 
        $0.1$ & $0.0994(1)$ & $0.060(3)$  & $0.868(5)$ & $0.54(3)$ & $0.988(6)$ \\ 
        $0.2$ & $0.0967(4)$ & $0.064(2)$  & $0.873(3)$ & $0.50(2)$ & $1.001(4)$ \\ 
        $0.3$ & $0.0848(1)$ & $0.060(6)$  & $0.87(1)$  & $0.50(3)$ & $0.99(1)$  \\ 
        $0.4$ & $0.0768(2)$ & $0.0626(9)$ & $0.878(3)$ & $0.50(4)$ & $1.004(4)$ \\ 
        $0.5$ & $0.0729(2)$ & $0.06(1)$   & $0.88(3)$  & $0.50(6)$ & $1.00(4)$  \\ 
        $0.6$ & $0.0706(4)$ & $0.060(2)$  & $0.877(7)$ & $0.50(5)$ & $0.997(8)$  \\ 
        $0.7$ & $0.0693(1)$ & $0.062(7)$  & $0.87(1)$  & $0.47(9)$ & $0.99(2)$  \\ 
        $0.8$ & $0.0685(3)$ & $0.062(3)$  & $0.885(7)$ & $0.50(2)$ & $1.009(8)$  \\ 
        $0.9$ & $0.0683(1)$ & $0.062(2)$  & $0.877(9)$ & $0.50(5)$ & $1.00(1)$  \\ 
        $1.0$ & $0.0678(3)$ & $0.064(2)$  & $0.878(4)$ & $0.50(1)$ & $1.008(4)$ \\ \hline
    \end{tabular}
    \setlength{\belowcaptionskip}{8pt}
   	\caption{\textbf{Volumetric finite-size scaling estimates of the critical behavior for different disorder levels.} The table reports the critical nonconformity $q_c$, the critical exponent ratios $\beta/\bar{\nu}$, $\gamma/\bar{\nu}$, $1/\bar{\nu}$, and the unitary relation $\upsilon$. Following the volumetric finite-size scaling framework proposed for complex interaction networks, the unitary relation is defined as $\upsilon \equiv 2\beta/\bar{\nu} + \gamma/\bar{\nu}$, which replaces the conventional hyperscaling relation when the correlation length scales with the system volume. The values obtained remain close to $\upsilon = 1$, supporting the volumetric scaling hypothesis for the majority-vote model on Gabriel graph networks.}
	\label{tab:criticalExpoforDiffs_volumetric}
\end{table*}

From the volumetric finite-size scaling relations of Eqs. (\ref{eq:scalinglnM}), (\ref{eq:scalinglnX}), and (\ref{eq:scalinglnU}), we rescale our measurements to obtain the universal functions

\begin{equation}\label{eq:scalinglnMtilde}
\widetilde{M} \left[u, N, s \right] = M(q, N, s)N^{\beta/\bar{\nu}},
\end{equation}

\begin{equation}\label{eq:scalinglnXtilde}
\widetilde{\chi} \left[u, N, s \right] = \chi(q, N , s)N^{-\gamma/\bar{\nu}},
\end{equation}

\begin{equation}\label{eq:scalinglnUtilde}
\widetilde{U} \left[u, N, s \right] = U(q, N, s),
\end{equation}
where $u = (q-q_{c})N^{1/\bar{\nu}}$. These functions enable data collapse, with rescaled observables obtained for different system sizes $N$ overlapping onto a single universal curve. In Fig. \ref{fig:dataCollapse-expoN}, panels (d), (e), and (f) present the corresponding volumetric data collapse obtained by rescaling the control parameter according to $(q-q_c)N^{1/\bar{\nu}}$. Each curve (color) comprises the network sizes $N = 1600, 3600, 6400, 10000$, and $14400$. For clarity, we shift each set of data symbols to avoid overlaps for different values of $s$. In Fig. \ref{fig:dataCollapse-expoN}(d), the rescaled magnetization data for each $s$ collapse onto a single universal curve, demonstrating that the collective consensus dynamics obey the same scaling function independently of system size $N$. Fig. \ref{fig:dataCollapse-expoN}(e) shows a similar collapse for the susceptibility, confirming that the amplification of collective opinion fluctuations near criticality is governed by universal scaling behavior. Fig. \ref{fig:dataCollapse-expoN}(f) presents the collapse of Binder’s cumulant, further validating the consistency of the estimated critical exponents and the scaling relations. The satisfactory quality of the data collapse across all observables indicates that the estimated critical exponents accurately capture the system's volumetric finite-size scaling behavior and consistently describe the universal properties of the consensus--dissent transition. These results also support the volumetric unitary relation $2\beta/\bar{\nu}+\gamma/\bar{\nu}=1$, confirming that the critical fluctuations scale consistently with the system volume in disordered Gabriel graph social networks.

\begin{figure*}[!ht]
	\centering
	\includegraphics[width=1.0\linewidth]{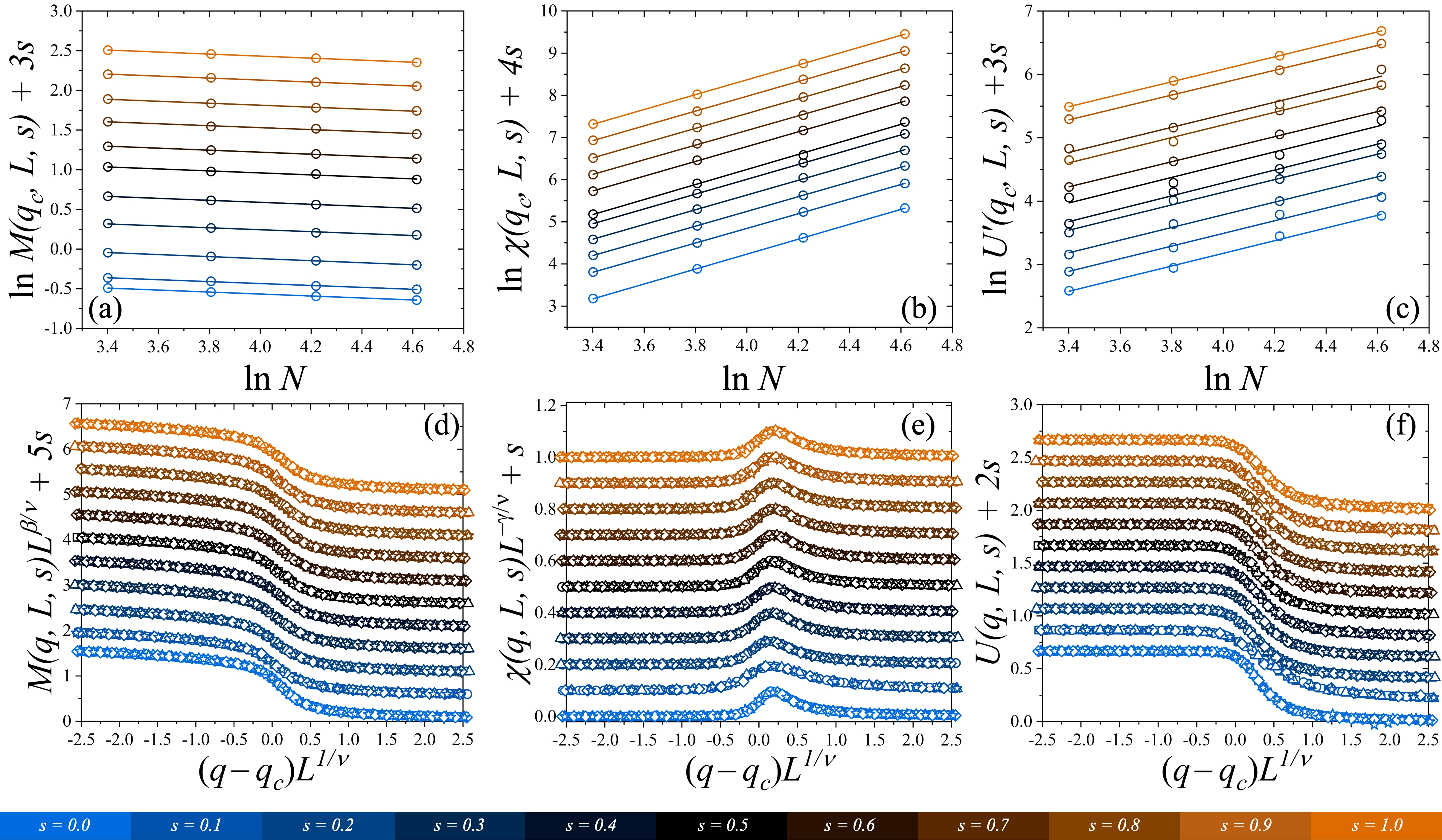}
	\caption{\textbf{Finite-size scaling and data collapse in linear representation.} Panels (a), (b) and (c) show log-log plot at the critical social temperature $q_c(s)$ as a function of the system linear size $L$: (a) average opinion $M(q_c, L,s)$, (b) susceptibility $\chi(q_c, L,s)$, and (c) derivative of Binder’s cumulant $U'(q_c, L,s)$. Panels (d), (e), and (f) show the data collapse obtained using the standard linear scaling, in which the control parameter is rescaled as $(q-q_c)L^{1/\nu}$, for (d) average opinion, (e) susceptibility, and (f) Binder's cumulant, in which each curve (color) comprises the linear network sizes $L = 40, 60, 80, 100$, and $120$. The hyperscaling relation demonstrates that the system has an effective dimension $d = 2$. The collapse of all observables onto single master curves across different disorder levels $s$ confirms that the majority-vote model on Gabriel graphs belongs to the two-dimensional Ising universality class. We shift each set of data symbols to avoid overlaps for different values of $s$ for clarity. The error bars are smaller than the symbol size.}
	\label{fig:dataCollapse-expoL}
\end{figure*}

We also perform an analogous finite-size scaling analysis using the linear size $L$, allowing a direct comparison with the conventional scaling framework of equilibrium critical phenomena. Analyzing the scaling behavior with respect to $L$ enables us to explore whether the nonequilibrium consensus--dissent transition preserves the hyperscaling properties and dimensionality expected for two-dimensional systems. This complementary investigation also provides an important consistency test between the volumetric scaling framework and the standard finite-size scaling formalism.

Fig. \ref{fig:dataCollapse-expoL} presents the finite-size scaling behavior and the corresponding data collapse obtained using the linear scaling representation. Panels (a), (b), and (c) show the log-log dependence of the critical observables evaluated at $q_c(s)$ as functions of $\ln L$. The approximately linear behavior observed for the magnetization, susceptibility, and derivative of Binder’s cumulant allows the estimation of the standard exponent ratios $\beta/\nu$, $\gamma/\nu$, and $1/\nu$, respectively. As in the volumetric representation, the magnetization decreases with increasing system size at criticality, while the susceptibility and cumulant derivative exhibit progressively stronger critical amplification. Table~\ref{tab:criticalExpoforDiffs_linear} presents the critical nonconformity levels and the corresponding linear critical exponent ratios for each disorder configuration. The estimated critical exponents remain compatible with a single linear universality class across the entire disorder range, and confirm that both representations encode similar critical behavior through the relation $N = L^d$ with effective dimension $d = 2$.

Fig. \ref{fig:dataCollapse-expoL}, panels (d), (e), and (f) present the corresponding data collapse obtained by rescaling the control parameter according to $(q-q_c)L^{1/\nu}$, in which each curve (color) comprises the network sizes $N = 1600, 3600, 6400, 10000$, and $14400$. The successful collapse of the magnetization, susceptibility, and Binder cumulant curves onto universal master curves demonstrates that the critical behavior is consistently described within the standard linear finite-size scaling framework. The results further support the standard hyperscaling relation $2\beta/\nu + \gamma/\nu = d$, yielding effective dimensionality values close to $d = 2$. This agreement demonstrates that the majority-vote dynamics on disordered Gabriel graph social networks preserve the universal critical behavior of two-dimensional systems despite the substantial geometric heterogeneity introduced by spatial disorder. These results reinforce the idea that disorder shifts the transition location without altering its universality class, agreeing with the Grinstein -- Jayaprakash hypothesis in which nonequilibrium models with short-range interactions, up-down symmetry, and single spin-flip dynamics fall into the Ising universality class at the respective dimension \cite{grinstein1985statistical}.

\begin{table*}[!ht]
    \centering
    \begin{tabular}{|c|c|c|c|c|c|}\hline
    \rowcolor{myorange!100}
        \textcolor{white}{$s$} & \textcolor{white}{$q_c$} & \textcolor{white}{$\beta / \nu$} & \textcolor{white}{$\gamma / \nu$} & \textcolor{white}{$1 / \nu$} & \textcolor{white}{$d$} \\ \hline\hline
        $0.0$ & $0.0750(1)$ & $0.1250(2)$ & $1.750(2)$ & $1.00(2)$ & $2.000(4)$ \\ 
        $0.1$ & $0.0994(1)$ & $0.120(6)$  & $1.736(1)$ & $1.08(6)$ & $1.976(1)$ \\ 
        $0.2$ & $0.0967(4)$ & $0.128(4)$  & $1.746(6)$ & $1.00(4)$ & $2.002(8)$ \\ 
        $0.3$ & $0.0848(1)$ & $0.120(1)$ & $1.74(2)$  & $1.00(6)$ & $1.98(2)$  \\ 
        $0.4$ & $0.0768(2)$ & $0.1252(2)$ & $1.756(6)$ & $1.00(8)$ & $2.008(8)$ \\ 
        $0.5$ & $0.0729(2)$ & $0.12(2)$   & $1.76(6)$  & $1.00(1)$ & $2.00(8)$  \\ 
        $0.6$ & $0.0706(4)$ & $0.120(4)$  & $1.754(1)$ & $1.00(1)$ & $1.994(2)$  \\ 
        $0.7$ & $0.0693(1)$ & $0.124(1)$ & $1.74(2)$  & $0.94(2)$ & $1.98(4)$  \\ 
        $0.8$ & $0.0685(3)$ & $0.124(6)$  & $1.770(1)$ & $1.00(4)$ & $2.018(2)$  \\ 
        $0.9$ & $0.0683(1)$ & $0.124(4)$  & $1.754(2)$ & $1.00(1)$ & $2.00(2)$  \\ 
        $1.0$ & $0.0678(3)$ & $0.128(4)$  & $1.756(8)$ & $1.00(2)$ & $2.016(8)$ \\ \hline
    \end{tabular}
    \setlength{\belowcaptionskip}{8pt}
    \caption{\textbf{Critical nonconformity levels and exponents for different disorder levels.} The estimated critical exponents remain consistent with the universality class of the two-dimensional Ising model, indicating that spatial disorder modifies the location of the consensus--dissent transition without changing the universal critical behavior of the majority-vote dynamics on Gabriel graph social networks.}
	\label{tab:criticalExpoforDiffs_linear}
\end{table*}

\section*{Discussion}

Introducing spatially constrained proximity interactions into the majority-vote opinion dynamics reveals how geometric organization directly influences collective social behavior. We incorporated spatial organization by constructing social interaction networks through Gabriel proximity graphs. Rather than assuming abstract or fully connected interactions, the model embeds agents in a two-dimensional geometric domain where local connectivity emerges directly from spatial constraints. As a consequence, the collective opinion behavior becomes intrinsically linked to the geometry of the interaction network. By introducing positional disorder into the spatial distribution of agents, we investigated how geometric heterogeneity modifies network organization and, consequently, the robustness of collective consensus formation under stochastic disagreement.

The results reveal that spatial disorder produces nontrivial structural transitions in the interaction topology. Weak geometric disorder enhances the average connectivity and clustering of the Gabriel networks by introducing additional local proximity connections, strengthening local reinforcement mechanisms, and stabilizing collective consensus. This effect manifests macroscopically as an increase in the critical nonconformity $q_c$, indicating that moderate positional heterogeneity makes the social system more robust against stochastic disagreement. However, at higher levels of positional disorder, the interaction structure progressively loses its local organization, reducing the effectiveness of collective coordination and shifting the transition toward lower values of $q_c$. The resulting phase diagram provides a direct relationship between spatial network organization and the stability of macroscopic social consensus.

The structural analysis reveals that the nonmonotonic behavior of the critical nonconformity emerges directly from disorder-induced modifications of the interaction topology. Weak positional disorder simultaneously enhances average connectivity and the clustering coefficient while partially reducing characteristic path lengths, strengthening local coordination, and facilitating the propagation of collective alignment throughout the network. Nevertheless, at higher levels of disorder, the progressive loss of geometric organization suppresses these effects, reducing the robustness of consensus despite the persistence of spatially constrained interactions. From a sociophysical perspective, these findings suggest that spatially constrained social systems may preserve robust universal collective behavior even when local interaction structures become strongly heterogeneous. The geometry of social interactions is essential in determining the stability of collective organization, and moderate structural disorder may even enhance consensus formation by improving local coordination and clustering. At the same time, excessive geometric disorder weakens the population's capacity to sustain global alignment under social nonconformity.

An important aspect of the present results is that the critical behavior remains compatible with the universality class of the two-dimensional Ising model despite the substantial geometric heterogeneity introduced by disorder. This contrasts with the behavior commonly observed in complex networks with nonlocal interactions and long-range shortcuts, where changes in universality properties frequently emerge. The successful scaling collapses of the magnetization, susceptibility, and Binder cumulant demonstrate that the nonequilibrium consensus--dissent transition preserves universal scaling behavior across the entire disorder range. These findings further support the Grinstein--Jayaprakash conjecture in spatially constrained nonequilibrium systems \cite{grinstein1985statistical}, indicating that symmetry and dimensionality remain the dominant factors governing the universal critical behavior even in the presence of substantial geometric disorder and intricate interaction topologies.

The volumetric finite-size scaling results indicate that the present system belongs to a volumetric counterpart of the two-dimensional Ising universality class. In the conventional linear formulation, the critical behavior of the two-dimensional Ising model is characterized by the exponent ratios $\beta/\nu = 1/8$, $\gamma/\nu = 7/4$, and $1/\nu = 1$, which satisfy the hyperscaling relation $2\beta/\nu + \gamma/\nu = d = 2$. Within the volumetric scaling framework, the volumetric exponent ratios naturally correspond to the conventional Ising exponents normalized by the spatial dimensionality $d$. Hence, for $d = 2$, we obtain the volumetric set $\beta/\bar{\nu} = 1/16$, $\gamma/\bar{\nu} = 7/8$, and $1/\bar \nu = 1/2$. The numerical estimates obtained throughout the disorder range remain compatible with these values, indicating that the volumetric scaling behavior preserves the universal properties of the two-dimensional Ising class while allowing the finite-size scaling description to be formulated directly in terms of the system volume $N$. This framework becomes especially meaningful for characterizing criticality in complex interaction topologies that do not naturally admit a well-defined linear scale, such as random graphs, scale-free networks, and, more generally, heterogeneous complex networks, where the notion of an underlying lattice structure or characteristic Euclidean separation between interacting elements is fundamentally absent. In this sense, the volumetric formulation provides a unified framework for comparing critical exponents between systems possessing a natural linear scale and systems whose topology prevents a direct geometric interpretation of length. From this perspective, we propose that the present model belongs to the volumetric counterpart of the two-dimensional Ising universality class characterized by the exponent ratios discussed above.

We associate the conservation of the effective dimensionality $d = 2$, as determined by the hyperscaling relation, even under heavily disordered configurations, with the planar and connected nature of Gabriel graph networks. Since the Gabriel criterion suppresses geometrically intersecting edges, the resulting topology remains spatially constrained and locally organized even under strong positional disorder. Consequently, the interaction structure preserves the dimension of the underlying two-dimensional embedding space. This feature distinguishes Gabriel graphs from other complex interaction topologies embedded in such spaces, where the introduction of nonlocal shortcuts significantly alters dimensionality \cite{2007HyunsukFiniteSize, 2011DaqingDimension, 2018SchrauthCoordination}. Previous results in Delaunay triangulations, relative-neighborhood graphs, and Voronoi-based structures further support the idea that planarity and connectedness emerge as crucial features in preserving two-dimensional critical behavior under topological disorder \cite{2018SchrauthCoordination, 2018BrendaVoronoi, 2019Alvesquasiperiodic, Schawe2017IsingProximity}. Such works establish important associations between planar and connected interaction structures and universality classes; however, the role of these geometric constraints in preserving the dimensionality of the embedding space remains largely unexplored. We conjecture that planar, spatially constrained, and connected networks embedded in two dimensions intrinsically exhibit hyperscaling with $d = 2$, even in the presence of strong topological disorder and heterogeneous connectivity patterns.

In light of these results, investigating other regular-to-heterogeneous planar networks may help determine whether spatially constrained topologies without geometrically crossing links retain the same dimensionality as the embedding space. Unlike interaction networks containing long-range shortcuts or nonlocal links, planar proximity structures preserve local spatial organization and constrain the propagation of correlations through the system. Consequently, the embedding geometry imposes robust constraints on universal properties of nonequilibrium collective dynamics, independent of substantial topological heterogeneity. Exploring whether this behavior extends to broader classes of planar networks and spin-based models may provide deeper insight into the relationship between topology and universality in spatially embedded complex systems.

\section*{Data availability}
Data will be made available on request.



\bibliography{referencesMVMGG.bib}

\section*{Author contributions}

F.A., G.G.P., and A.L.M.V. conceived and implemented the models and numerical experiments, performed the simulations, and produced the figures. A.J.F.S., P.R.A.C., and A.L.M.V. supervised the project and contributed to the analysis and interpretation of the results. F.A. wrote the first version of the manuscript, and A.L.M.V. revised and improved the final text. All authors reviewed and approved the final manuscript.

\section*{Competing interests}
The authors declare no competing interests.\\

\noindent \textbf{Correspondence} and requests for materials should be addressed to Andr\'e L. M. Vilela (email: \href{mailto:andre.vilela@upe.br}{andre.vilela@upe.br}).

\section*{Acknowledgments}
The authors acknowledge financial support from Brazilian institutions and funding agents UPE, FACEPE (APQ-1129-1.05/24), CAPES, and CNPq (306336/2025-1, 301795/2022-3). We acknowledge support from the Center and Laboratory for Simulation of Complex Systems, Recife, Brazil, for providing the computational resources and infrastructure necessary for the numerical simulations. We used OpenAI's ChatGPT to assist with language refinement and manuscript editing. All scientific content, analysis, and interpretation were solely developed by the authors.

\end{document}